# Teaching the Doppler Effect in Astrophysics

*Stephen W. Hughes[1], Michael Cowley[2,3]*

[1]*Department of Chemistry, Physics and Mechanical Engineering,*
*Queensland University of Technology, Gardens Point Campus,*
*Brisbane, Queensland 4001, Australia*
*Phone: +617 3285 3087*
*Email: sw.hughes@qut.edu.au*
[2]*Department of Physics & Astronomy,*
*Macquarie University, Sydney, NSW 2109, Australia*
[3]*Australian Astronomical Observatory,*
*PO Box 915, North Ryde, NSW 1670, Australia*

**Abstract**
The Doppler effect is a shift in the frequency of waves emitted from an object moving relative to the observer. By observing and analysing the Doppler shift in electromagnetic waves from astronomical objects, astronomers gain greater insight into the structure and operation of our universe. In this paper, a simple technique is described for teaching the basics of the Doppler effect to undergraduate astrophysics students using acoustic waves. An advantage of the technique is that it produces a visual representation of the acoustic Doppler shift. The equipment comprises a 40 kHz acoustic transmitter and a microphone. The sound is bounced off a computer fan and the signal collected by a DrDAQ ADC and processed by a spectrum analyser. Widening of the spectrum is observed as the fan power supply potential is increased from 4 to 12 V.

**Introduction**
The Doppler effect is used in many areas of astrophysics. For example, Doppler measurements can be used to probe the gravitational field strength above the surface of white dwarfs [1], map the velocity distribution of stars in the Milky Way [2], and model the internal structure of the Sun and other stars by observing ripples generated by acoustic waves on the stellar surface [3]. Other applications include Doppler spectroscopy, which can reveal the subtle motion caused by an exoplanet [4], Zeeman Doppler Imaging (ZDI) to map the magnetic fields of stars by observing the Zeeman splitting of spectral lines [5], and redshift measurements to measure the recessional velocity of distant galaxies [6] and velocity distribution of material in supernova remnants [7]

There are a number of articles in the literature that demonstrate various approaches to testing and teaching the Doppler effect. For example, [8] use a water ripple tank to analyse straight waves, while [9] use a smart phone in a novel experiment to measure the Doppler effect of linear motions. Here, we describe a simple way of visually demonstrating the acoustic Doppler effect of a circular motion system (i.e. a rotating fan) in the teaching laboratory.

Although in astrophysics the Doppler effect predominantly involves electromagnetic waves, the principle demonstrated using acoustic waves is the same. In astrophysics relativistic effects often come into play which is not applicable to acoustic Doppler, and so the experiment in this paper elucidates the principle of electromagnetic Doppler effect when the velocity is much less than the speed of light.

The Doppler shift, or redshift ($z$), is defined as

$$z = \frac{\lambda - \lambda_0}{\lambda_0} = \frac{\Delta\lambda}{\lambda_0}$$

The change in wavelength ($\Delta\lambda$) is related to the speed of an object. An object with radial velocity ($v$), i.e. a velocity towards or away from the Earth, moves a distance $vT$ in the period ($T$) of the wave. The wavelength is stretched or compressed by this amount depending on whether the component of the radial velocity is towards or away from the observer.

$$\Delta\lambda = vT = \frac{v}{f} = \frac{v\lambda}{c} \Rightarrow \frac{\Delta\lambda}{\lambda} = \frac{v}{c}$$

$$z = \frac{\Delta\lambda}{\lambda} = \frac{v}{c}$$

$$\frac{\Delta\lambda}{\lambda} = \frac{\Delta f}{f}$$

$$\frac{\Delta f}{f} = \frac{v}{c} \therefore \Delta f = f\frac{v}{c}$$

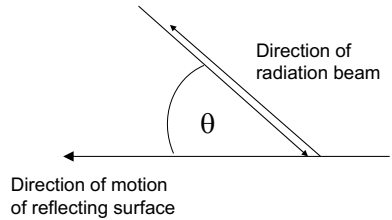

Figure 1. Doppler shift is dependent on the cosine of the angle between the direction of motion and radiation beam.

If the motion of the radiation source has a non-radial component the frequency shift scales with the cosine of the angle between the direction of motion and the line joining the Earth and object. The Doppler shift equation for an emitted wave is:

$$\Delta f = f \frac{v}{c} \cos\theta$$

where $\Delta f$ is the Doppler shift frequency, $f$ is the carrier frequency, $v$ is the velocity of object, $c$ is the velocity of radiation, $\theta$ is the angle between motion of object and direction of radiation (figure 1).

From an educational point of view it is interesting to compare this equation with the ultrasound Doppler equation, which has an additional factor of 2 because there is a Doppler shift when sound is absorbed and reflected.

$$\Delta f = 2f \frac{v}{c} \cos\theta$$

The factor of 2 is not required in astrophysics although it is required in fan Doppler measurements session since ultrasound is reflected off a moving object. A more realistic astrophysical situation would be to attach the ultrasound source to the fan blades. However, it can be appreciated that attaching an ultrasonic transmitter to moving fan blades would be difficult.

In this experiment a fan with a 55 mm radius was used with a nominal rotation rate of 3000 rpm (50 Hz). At 3000 rpm the speed of the edge of the fan is:

$$v = \pi d f \cong 3.14 \times 0.11 \times 50 = 17.3 \text{ ms}^{-1}$$

where

$v$ = velocity of the tips of the fan blades
$d$ = diameter of fan blades
$f$ = rotation frequency in rotations per second.

For a peak velocity of 17.3 ms$^{-1}$, the Doppler shift frequency is:

$$\Delta f = 2f \frac{v}{c} \cos \theta = \frac{2 \times 4 \times 10^4 \times 17.3}{3.3 \times 10^2} = 4.2 \text{ kHz}$$

When the fan blade is moving away from the source, the redshifted signal is 40 kHz + 4.2 kHz = 44.2 kHz. When the fan blade is moving towards the source, the blue-shifted signal is 40 kHz − 4.2 kHz = 35.8 kHz. A shift of this magnitude is easily seen in the spectrum. In practice the shift will be less since most of the fan is moving slower since the blades are between 2 and 5.5 cm from the centre, and the angle between the sound waves and the plane of rotation is greater than zero degrees.

The experiment described here is analogous to the use of the Doppler Effect in medical ultrasound where the change in frequency of sound with a frequency of several MHz reflected off blood cells falls within the audio frequency range. In this experiment the Doppler shift from the blades of a spinning computer fan is a few kHz and therefore large enough to see when 40 kHz sound is reflected off the fan blades.

**Method**
The equipment comprises two ultrasonic transducers, a transmitter and a receiver that operate optimally at 40 kHz. The transmitter and receiver are projected towards the fan – the angle between the two is not important since the aim of this experiment is to just demonstrate widening of the spectra of the reflected acoustic wave.

The kit was obtained from a local electronics store (Jaycar) for about A$30. The output of the receiver was connected to a Picotech DrDAQ analogue to digital converter (ADC) set to a sampling frequency of 100 kHz, i.e. a sampling rate of 100 000 samples per second. The ADC has 8-bit resolution and is connected to a PC via a USB connector. A photo of the complete equipment is shown in figure 2 and a schematic diagram in figure 3.

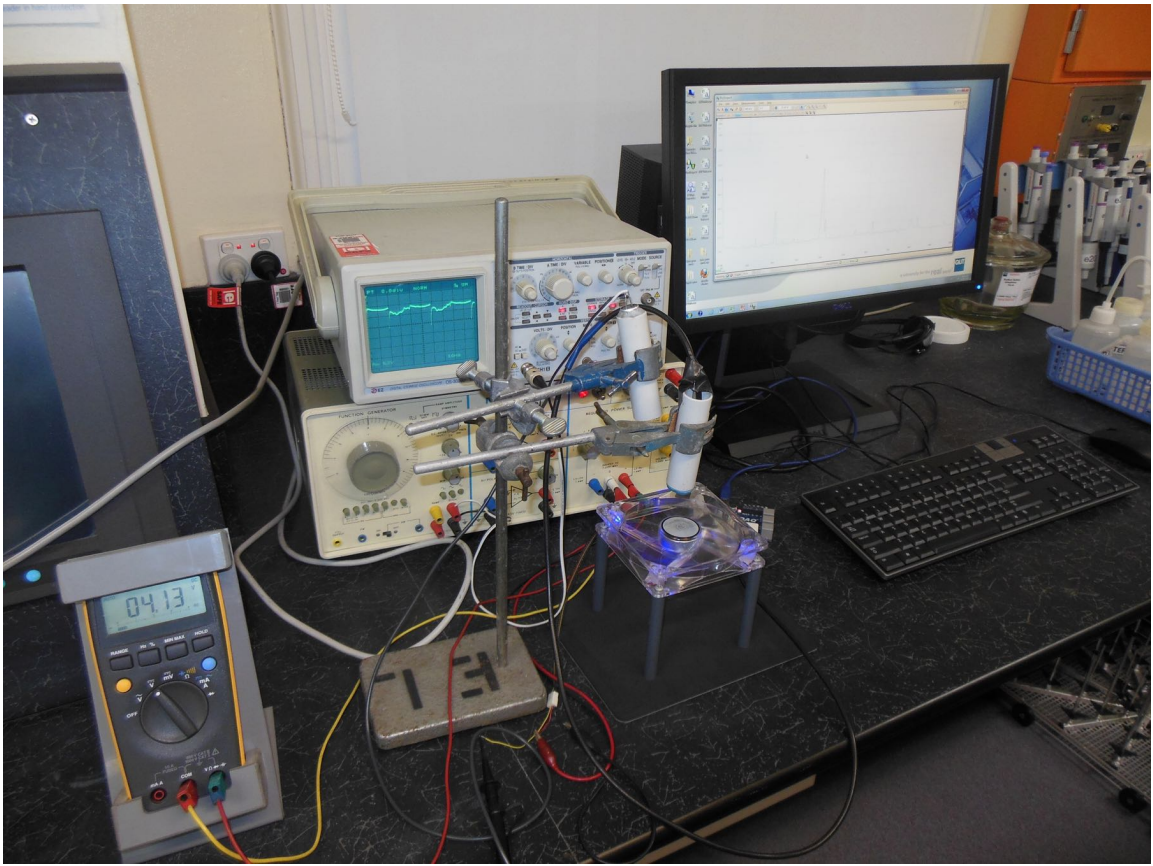

Figure 2. The complete equipment. The rpm trace can just be seen on the CRO on top of the power supply.

The fan motor has a three-wire connection – red for the power supply, black 0 V and yellow for the revolutions per minutes (rpm) output. One pulse is produced for each revolution.

The red and black wires are attached to a 0-15 VDC power supply and the rpm wire to a CRO probe. In the student experiment the rpm trace is displayed on a conventional CRO since the Dr.DAQ only has a single CRO input. The ultrasound transmitter and receiver were placed in the end of plastic tubes and held in place with Blu-Tak. These are oriented facing down to the fan. The fan is placed above a rubber mouse pad to reduce acoustic reflections.

The transmitter is connected to a signal generator integral to the lab power supply and the receiver connected directly to the DrDAQ CRO probe connection. The CRO probe of a digital oscilloscope is connected to the yellow rpm wire. (Although this output is called the rpm, one pulse is generated per revolution, therefore the inverse of the temporal separation of pulses gives the rotation rate in revolutions per second).

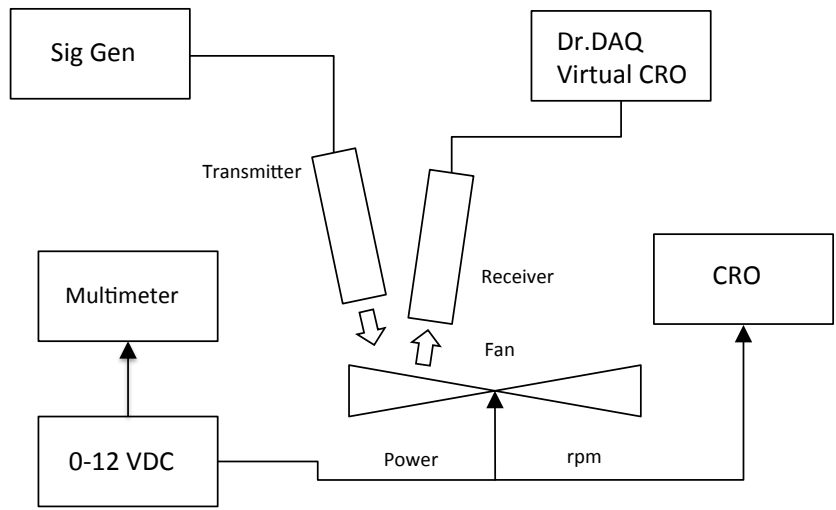

Figure 3. Schematic diagram of the equipment.

When the PicoScope program is run, by default a blue trace appears in the scope window – this is the input from the microphone built into the DrDAQ. (This can be verified by clapping). The audio input is switched off and the CRO input displayed on the screen. The spectrum mode is then selected by clicking on the spectrum icon. A noisy spectrum appears but the noise can be reduced by setting the average display mode to 512 bins.

To collect data for this paper, initially the fan was off and a sharp peak is seen at 40 kHz. The fan was switched on and the spectrum collected for power supply voltages of 4, 6, 8, 10 and 12 V. About 20 seconds is required for the spectrum to stabilise for each voltage. After freezing the trace, the data was stored in comma separated variable format (csv), enabling the data to be directly imported into Excel. The data is stored as a column of time values with a corresponding CRO voltage. The time separation between the rpm pulses is measured on the digital CRO for each fan power supply voltage.

There are various ways of analysing the data. In this case each spectrum was assumed to approximate a triangle. The area (A) of a triangle is given by half the product of the base (b) and height (h), $A = bh/2$. The area of each spectrum was calculated and the base of the equivalent triangle was found from $b = 2A/h$. For the data shown in this experiment the spectra were integrated between 37 and 43 kHz.

**Results**

Table 1 shows the basic data. Figure 4 shows a plot of the spectrum for fan potentials of 0, 8 and 12 V. The plot shows that the width of the spectrum increases with fan rotation speed. Figure 5 shows the width of each spectra for 0, 4, 6, 8, 10 and 12 V.

Table 1. Basic experimental data.

| Fan potential (V) | Fan rotation period (ms) | Fan frequency (Hz) | Doppler shift (kHz) |
|---|---|---|---|
| 0 | 0 | 0 | 2.06 |
| 4 | 41 | 24.39 | 2.5 |
| 6 | 29 | 34.48 | 2.81 |
| 8 | 21 | 47.62 | 3.18 |
| 9 | 18 | 55.56 | 3.43 |
| 12 | 16 | 62.50 | 3.52 |

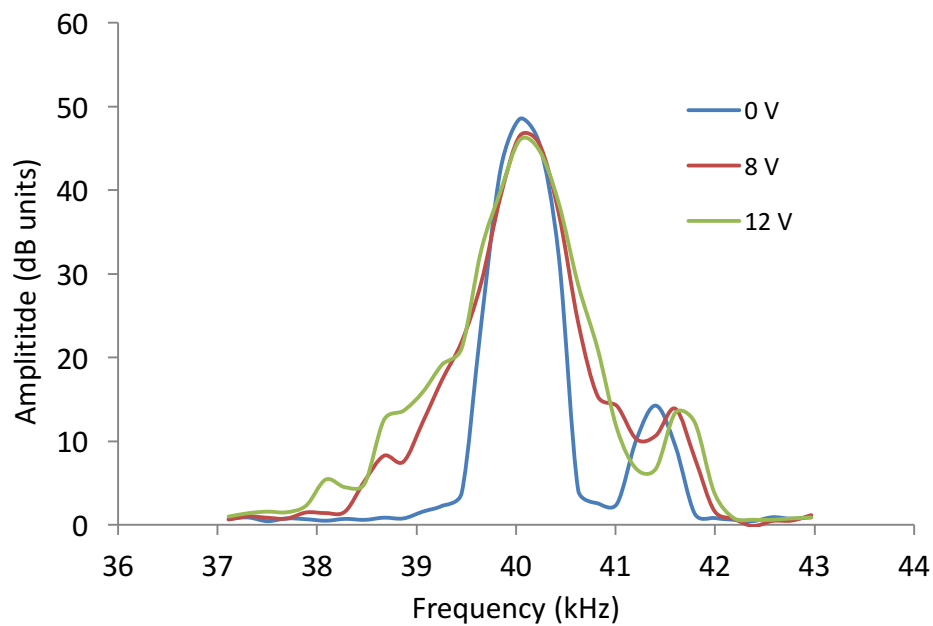

Figure 4. Plot of the spectra for power supply potentials of 0, 8 and 12 V, showing widening of the spectrum with increasing fan rotation speed.

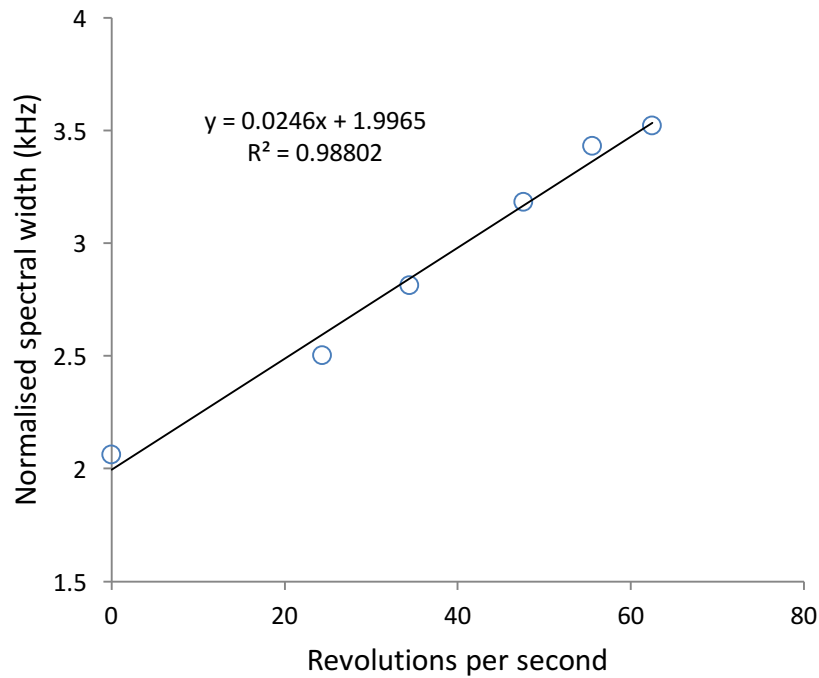

Figure 5. Plot of spectra widths versus fan frequency.

**Discussion**

The results show that when the fan power supply potential is increased, there is a noticeable widening of the spectra around the central frequency of 40 kHz. The plot of the spectrum width versus the fan rotation frequency is a reasonable straight line as expected.

The experiment described in this paper is fairly easy to set up. Virtually all the required equipment is available in a university physics department – i.e. DC power supply, signal generator, multimeter, CRO. The only additional pieces of equipment required are the Dr.DAQ ADC, computer fan and ultrasound kit. Although the experiment described here is used to teach second year astrophysics, the experiment could also be useful for teaching the fundamentals of ultrasound in a medical physics course.

The experiment could be performed with a sound source with a frequency in the acoustic range in which case the Doppler shift would be audible as well as graphical. The reason for using 40 kHz in this experiment was due to the availability of a transmitter/receiver pair.

**References**

[1] Adams W 1925 The relativity displacement of the spectral lines in the companion of Sirius *PNAS* **11** 382-287

[2] Heath Jones D, Saunders W, Colless M, Read, M A, Parker Q A, Watson F G, Campbell L A, Burkey D, Muach T, Moore L, Hartley M, Cass P, James D, Russell K, Fiegert K, Dawe J, Huchra J, jarrett T, Lahav O, Lucey J, Mamon G A, Proust D, Sadler AE M, Wakamatsu K 2004 The 6dF galaxy survey: samples, observational techniques and first data release *MNRAS* **for** 747-763

[3] Tomczyk S, Streander K, Card G, Elmore D, Hull H, Cacciani A 1995 An instrument to observe low-degree solar oscillations *Sol. Phys* **159** 1-21, 1995

[4] Mayor M, Queloz D 1995 A Jupiter-mass company on to a solar-type star *Nature* **378** 355-359

[5] Semel M, Donati J -F, Rees D E 1993 Zeeman-Doppler imaging of active stars III. Instrumental and technical considerations *Astron. Astrophys.* **278** 231-237

[6] Hubble E 1929 A relation between distance and radial velocity among extra-galactic nebulae *PNAS* **15** 168-173

[7] Hayato A, Yamaguchi H, Tamagawa T, Katsuda S, Hwang U, Hughes P, Ozawa M, bamaba A, Kinugasa K, Yukikatsu T, Furuzawa A, Kunieda H, Makishimia K 2010 Expansion velocity of ejecta in Tycho's supernonva remnant measured by Doppler broadened x-ray line emission, *ApJ* **725** 894-903

[8] D'Anna M and Corridoni T 2016 An old experiment revisited: the Doppler effect in a ripple tank *Eur. J. Phys.* **37** 045703

[9] Gómez-Tejedor J A, Castro-Palacio J C and Monsoriu J A 2014 The acoustic Doppler effect applied to the study of linear motions *Eur. J. Phys.* **35** 025006